\documentclass[12pt]{article}
\usepackage{amssymb,amsmath,epsfig}

\begin{document}

\title{\bf Study of Inflationary Generalized Cosmic Chaplygin Gas for Standard and Tachyon Scalar Fields}
\author{M. Sharif \thanks {msharif.math@pu.edu.pk} and Rabia Saleem
\thanks{rabiasaleem1988@yahoo.com}\\
Department of Mathematics, University of the Punjab,\\
Quaid-e-Azam Campus, Lahore-54590, Pakistan.}

\date{}
\maketitle

\begin{abstract}
We consider an inflationary universe model in the context of
generalized cosmic Chaplygin gas by taking matter field as standard
and tachyon scalar fields. We evaluate the corresponding scalar
fields and scalar potentials during intermediate and logamediate
inflationary regimes by modifying the first Friedmann equation. In
each case, we evaluate the number of e-folds, scalar as well as
tensor power spectra, scalar spectral index and important
observational parameter, i.e., tensor-scalar ratio in terms of
inflatons. The graphical behavior of this parameter shows that the
model remains incompatible with WMAP7 and Planck observational data
in each case.
\end{abstract}

\section{Introduction}

A combination of different cosmic probes like type Ia supernova, the
large scale structure (LSS), cosmic microwave background (CMB) and
WMAP confirmed that our universe is experiencing accelerating
expansion \cite{1}. Little is known about the origin of this cosmic
stage which may be due to dark energy (DE) (with large negative
pressure). It fills two-third of the whole cosmic energy and
remaining portion is almost occupied by the dark matter (DM). A tiny
constant $\Lambda$ is the simplest identification of DE which
suffers with fine-tuning and cosmic coincidence issues. The
dynamical nature of DE is divided into scalar field models
(quintessence, phantom, k-essence etc.) \cite{3} and interacting DE
models (Chaplygin gas (CG), holographic DE, Ricci DE etc.) \cite{4}.

Chaplygin gas (a unification of DE and DM) is considered to be an
interesting alternative description of accelerating expansion. It
has negative pressure obeying equation of state (EoS)
$P=-\frac{A}{\rho},~A>0$ and positive speed of sound which is a
powerful tool to discriminate between various DE models. The
velocity of sound approaches to the velocity of light for late time
while negligibly small for early times. The energy density of CG
smoothly varies from matter dominated era to a constant point, i.e.,
$\Lambda$ cold DM ($\Lambda$CDM) in the future universe \cite{5}.
Many people carried out cosmology of different models of CG like
generalized CG (GCG) \cite{6}, modified CG (MCG) \cite{7} and
generalized cosmic CG (GCCG) \cite{34} etc. Kamenshchik et al.
\cite{a} considered FRW universe composed of CG and showed that
resulting evolution of the universe is in agreement with the current
observation of cosmic acceleration.

Recently, a great amount of work has been done in investigating the
inflationary universe model with a tachyon field. This field might
be responsible for cosmological inflation in the early evolution of
the universe due to tachyon condensation near the top of the
effective scalar potential \cite{11}, which could also add some new
form of cosmological DM at late times \cite{12}. Gibbons \cite{13}
was the first who studied cosmological implications of this rolling
tachyon. It is quite natural to consider some scenarios in which
inflation is driven by the rolling tachyon. The CG emerges as an
effective fluid of generalized d-brane in a (d+1, 1) spacetime,
where the action can be written as a generalized Born-Infeld action
\cite{14}. These models (CG and tachyon) have extensively been
studied in the literature \cite{15}. In Chaplygin inspired
inflationary universe model, the standard inflaton field usually
drives inflation where the energy density can be extrapolated for
obtaining a successful inflation period \cite{16}. Del Campo and
Herrera \cite{17} studied warm-Chaplygin and tachyon-Chaplygin
inflationary universe model. Monerat et al. \cite{18} explored
dynamics of the early universe and initial conditions for an
inflationary model with radiation and CG.

The standard cosmology explains observations of CMB radiations in an
elegant way but early phase of the universe is still facing some
long-standing issues like horizon problem, flatness, numerical
density of monopoles and the origin of fluctuations \cite{19}. The
inflationary models present better description of the early universe
which also provide the most compelling solution of these problems.
Inflation can provide an elegant mechanism to explain causal
interpretation of the origin of the observed anisotropy of CMB and
inhomogeneity for structure formation. Scalar field models composed
of kinetic and potential terms coupled to gravity produce dynamical
framework and act as a source for inflation. These models have
ability to interpret the distribution of LSS and observed anisotropy
of CMB radiations comprehensively in inflationary era \cite{20}.

Inflationary era is divided into slow-roll and reheating epochs.
During slow-roll approximation, the universe inflates as the
interactions between inflatons and other fields become negligibly
small and potential energy dominates the kinetic energy. After this
period, the universe enters into last stage of inflation, i.e.,
reheating era in which kinetic and potential energies are
comparable. Here the inflaton starts to oscillate around the minimum
of its potential while losing its energy to massless particles.
Inflationary model is usually discussed in intermediate and
logamediate scenarios.

During intermediate era, the universe expands at the rate slower
than the standard de Sitter inflation while faster than power-law
inflation \cite{23}. Setare and Kamali \cite{30} have discussed warm
vector inflation in this scenario for FRW model and proved that the
results are compatible with WMAP7 data \cite{29}. The same authors
\cite{32} also dealt with warm inflation using gauge fields in
intermediate as well as logamediate scenarios. In a recent paper
\cite{32*}, we have studied warm vector inflation in locally
rotationally symmetric Bianchi type I universe model and verified
its compatibility with WMAP7 data.

The study of inflationary epoch with intermediate and logamediate
scale factors lead to over-lasting forms of the potential which
agree with tachyon potential properties. Moreover, the study of warm
inflation as a mechanism gives an end for standard and tachyon
inflation. This motivated us to consider inflationary model with
these two potentials. Recently, Herrera et al. \cite{33} studied
intermediate GCG inflationary universe model with standard as well
as tachyon scalar fields and checked its compatibility with WMAP7
data. Since GCCG is less constrained as compared to MCG and GCG and
is capable of adapting itself to any domain of cosmology depending
upon the choice of parameters. Thus it has more universal character
and the big-rip singularity can easily be avoided in this model.
These generalizations of CG can lead to significant changes in the
early universe. It would be interesting to check the behavior of
inflationary universe with GCCG using standard and tachyon scalar
fields during intermediate as well as logamediate epochs. This work
can recover all the previous existing models of CG.

The paper is arranged in the following format. In the next section,
we modify the first Friedmann equation and find solutions of
standard and tachyon scalar fields as well as their corresponding
potentials. We also provide the slow-roll parameters, number of
e-folds, scalar and tensor power spectra, scalar spectral index and
tensor-scalar ratio. In section \textbf{3}, we develop our model in
intermediate and logamediate inflation with both types of scalar
fields. We conclude our discussion in the last section.

\section{Inflation with Standard and Tachyon Scalar Fields}

In this section, we modify the first Friedmann equation in the
context of GCCG inflationary universe model. We choose standard and
tachyon scalar fields as matter content of this universe and
calculate both scalar fields and their corresponding potentials. We
also formulate some important perturbed parameters.

Gonz\'{a}lez-Diaz \cite{34} introduced GCCG model in such a way that
the resulting models can be made stable and physical even when the
vacuum fluid satisfies the phantom energy condition. It has the
following exotic EoS
\begin{equation}\label{1}
P=-\rho^{-\alpha}\left[C+(\rho^{1+\alpha}-C)^{-\omega}\right],
\end{equation}
where $C=\frac{A}{1+\omega}-1,~A$ is either positive or negative
constant, $\alpha$ is any positive constant and $-l<\omega<0$,\quad
$l>1$. This EoS reduces to GCG model as $\omega\rightarrow0$. The
corresponding energy density is obtained by integrating the energy
conservation equation of the GCCG as follows
\begin{equation}\label{2}
\rho=\left[C+\left(1+\frac{B}{a^{3(1+\alpha)(1+\omega)}}\right)
^{\frac{1}{1+\omega}}\right]^{\frac{1}{1+\alpha}},
\end{equation}
with scale factor $a$ and $B$ is the integration constant. The
gravity dynamics during inflation leads to modify the first
Friedmann equation as \cite{14}
\begin{equation}\label{3}
H^2=\frac{\kappa}{3(1+\omega)}\left[C+
\rho^{(1+\alpha)(1+\omega)}_{\phi}\right]^{\frac{1}{1+\alpha}},
\end{equation}
where $\kappa=\frac{8\pi}{m^{2}_{p}},~m_{p}$ is the reduced Planck
mass and $\rho_{\phi}$ is the energy density of the scalar field.
This modification is dubbed as Chaplygin inspired inflation.

We take two types of scalar fields for $\rho_{\phi}$, i.e., standard
and tachyon scalar fields. The energy conservation of a scalar field
is
\begin{equation}\label{4}
\dot{\rho}_{\phi}+3H(\rho_{\phi}+P_{\phi})=0,
\end{equation}
where the associated standard energy density and pressure are given
as
\begin{equation*}
\rho_{\phi}=\frac{\dot{\phi}^2}{2}+V(\phi), \quad
P_{\phi}=\frac{\dot{\phi}^2}{2}-V(\phi).
\end{equation*}
Using $\rho_{\phi}$ and $P_{\phi}$, the above equation is equivalent
to the equation of motion of the standard scalar field as follows
\begin{equation}\label{5}
\ddot{\phi}+3H\dot{\phi}+V^{\prime}(\phi)=0,
\end{equation}
where $\prime$ denotes derivative with respect to $\phi$. Equations
(\ref{3}) and (\ref{4}) yield
\begin{equation}\label{6}
\dot{\phi}^{2}=-\left(\frac{2\dot{H}}{\kappa}\right)
\left(\frac{3H^2}{\kappa}\right)^{\alpha}\left(\frac{3H^2}
{\kappa}\right)^{(1+\alpha)\psi}\left[1-\frac{A}{(1+\omega)}
\left(\frac{\kappa}{3H^2}\right)^{1+\alpha}\right]^{\psi},
\end{equation}
where $\psi=\left[\frac{1}{(1+\alpha)(1+\omega)}-1\right]$. The
scalar potential is obtained by substituting $\rho_{\phi}$ from
(\ref{3}) and $\dot{\phi}^{2}$ from (\ref{6}) in the formula for
energy density of standard scalar field as follows
\begin{eqnarray}\nonumber
V(t)&=&(1+\omega)\left[\left(\frac{3H^2}{\kappa}\right)
^{1+\alpha}-\frac{A}{(1+\omega)}\right]^{\psi+1}
+\left[\frac{\dot{H}}{\kappa}\left(\frac{3H^2}{\kappa}\right)
^{\alpha+(1+\alpha)\psi}\right.
\\\label{7}&\times&\left.\left[1-\frac{A}
{(1+\omega)}\left(\frac{\kappa}{3H^2}\right)^{1+\alpha}\right]
^{\psi}\right].
\end{eqnarray}
The above two solutions reduce to typical standard inflation for
$\alpha,~A,~\omega\rightarrow0$, pure CG model for
$\alpha,~\omega\rightarrow0$ and GCG model for $\omega\rightarrow0$
\cite{25}.

The energy density and pressure of tachyon field are
\begin{equation}\label{1*}
\rho_{\phi}=\frac{V(\phi)}{\sqrt{1-{\dot{\phi}}^{2}}},\quad
P_{\phi}=V(\phi)\sqrt{1-{\dot{\phi}}^{2}}.
\end{equation}
Using Eq.(\ref{4}), we obtain the corresponding equation of motion
\begin{equation}\label{2*}
\frac{\ddot{\phi}}{1-\dot{\phi}^{2}}+3H\dot{\phi}+\frac{V^{\prime}(\phi)}{V(\phi)}=0.
\end{equation}
Equations (\ref{3}) and (\ref{2*}) provide the time derivative of
tachyon field as follows
\begin{equation}\label{3*}
\dot{\phi}^{2}=-\left(\frac{2\dot{H}}{\kappa}\right)
\left(\frac{3H^2}{\kappa}\right)^{\alpha}\frac{1}{(1+\omega)}\left[
\left(\frac{3H^2}{\kappa}\right)^{1+\alpha}-\frac{A}{1+\omega}\right]^{-1}.
\end{equation}
Using above equation with (\ref{3}) in $\rho_{\phi}$ given in
(\ref{1*}), we have tachyon potential
\begin{eqnarray}\nonumber
V(t)&=&(1+\omega)^{\frac{1}{2}}\sqrt{1+\frac{2\dot{H}}{\kappa}
\left(\frac{3H^2}{\kappa}\right)^{\alpha}\frac{1}{(1+\omega)}\left[
\left(\frac{3H^2}{\kappa}\right)^{1+\alpha}-\frac{A}{1+\omega}\right]^{-1}}
\\\label{4*}&\times&\left[\left(\frac{3H^2}{\kappa}\right)
^{1+\alpha}-\frac{A}{(1+\omega)}\right]^{\psi+1}.
\end{eqnarray}
The dimensionless slow-roll parameters $\epsilon,\eta$ and number of
e-folds are defined as
\begin{equation}\label{13*}
\epsilon=-\frac{\dot{H}}{H^2},\quad\eta=-\frac{\ddot{H}}{H\dot{H}},\quad
N=A\int^{t_{2}}_{t_{1}}Hdt;\quad A>0,
\end{equation}
where $t_{1}$ and $t_{2}$ being the starting and ending cosmic time
of inflationary era.

Now we define scalar and tensor power spectra for GCCG inflationary
model with standard and tachyon scalar fields. The power spectrum as
a function of wave number $(k)$ is the basic tool to quantify
fluctuation's variance produced by inflatons. In order to calculate
scalar perturbation, a gauge invariant quantity,
$\zeta=H+\frac{\delta\rho}{\dot{\rho}}$, is introduced \cite{36}.
This quantity almost remains constant on super-horizon scales but
reduces to curvature perturbation on a slice of uniform density.
This fundamental characteristic is a consequence of stress-energy
conservation and independent of gravitational dynamics which keeps
it unchanged in Chaplygin inflationary model \cite{37}. Thus the
power spectrum corresponds to curvature spectrum and can be written
as $\mathcal{P_{R}}=\langle\zeta^2\rangle$ \cite{33}. Since the
curvature perturbations act as comoving curvature perturbation on
the slices of uniform density, so for spatially flat gauge fields,
we have \cite{38}
\begin{equation}\label{7*}
\mathcal{P_{R}}\simeq H^2\frac{(\delta
\phi)^{2}}{(\dot{\phi})^{2}};\quad|\delta
\phi|=\frac{H}{2\pi}\quad\Rightarrow\quad
\mathcal{P_{R}}\simeq\frac{H^4}{4\pi^2\dot{\phi}^{2}}.
\end{equation}
The scalar power spectrum for tachyon field using slow-roll
approximation $((\dot{\phi})^2<<V(\phi))$ has the form \cite{39}
\begin{equation}\label{11*}
\mathcal{P_{R}}\simeq\left(\frac{H^2}{2\pi\dot{\phi}}\right)^2\frac{1}{Z_{s}};
\quad Z_{s}=V(1-(\dot{\phi})^2)^{-\frac{3}{2}}\approx V(\phi).
\end{equation}
The tensor perturbation generating gravitational waves and scalar
spectral index, $n_{s}$ are defined as
\begin{equation}\label{8*}
\mathcal{P}_{g}=8\kappa\left(\frac{H}{2\pi}\right)^{2},\quad
n_{s}-1=-\frac{d\ln \mathcal{P_{R}}(k)}{d\ln k}.
\end{equation}

The tensor-scalar ratio (an observational quantity) for both
standard and tachyon scalar fields, respectively, is
\begin{equation}\label{10*}
r=\frac{\mathcal{P}_{g}}{\mathcal{P_{R}}}=8\kappa\left(\frac{\dot{\phi}}{H}\right)^{2},\quad
r=8\kappa\left(\frac{\dot{\phi}}{H}\right)^{2}V.
\end{equation}
According to the observations of WMAP+BAO (baryon acoustic
oscillations)+SN, the scalar spectral index and perturbed scalar
power spectrum are constrained to $0.96\leq n_{s}\leq1.002$ (95\%
C.L.) and
$\mathcal{P_{R}}(k_{0}=0.002Mpc^{-1})=(2.445\pm0.096)\times10^{-9}$,
respectively \cite{20} while tensor power spectrum cannot be
constrained directly. In this context, physical acceptable range of
tensor-scalar ratio is determined, i.e., $r<0.36$ (95\% C.L.) which
represents the expanding universe.

\section{Intermediate and Logamediate Inflation}

Here, GCCG inflationary universe model is developed in intermediate
and logamediate eras using standard and tachyon scalar fields. We
reconstruct solutions of both fields, their potentials and perturbed
parameters (found in the above section) during these two scenarios.

\subsection{Standard Scalar Field}

First, we take standard scalar field as matter content of the
inflationary universe and discuss in intermediate as well as
logamediate scenarios.

\subsubsection{Intermediate Inflation}

This era is motivated by string/M theory and is one of the exact
solutions of the inflationary cosmology. The 4-dimensional Gauss
Bonnet interaction with dynamical dilatonic scalar coupling leads to
an intermediate form of the scale factor \cite{24}
\begin{equation}\label{a}
a(t)=a_{0}\exp(At^{f}),\quad A>0,~0<f<1,
\end{equation}
where $a_{0}$ is the value of scale factor at $t=0$. Using
Eq.(\ref{a}) in (\ref{6}), we obtain the following solution of
standard scalar field $\phi$
\begin{equation}\label{8}
\phi(t)-\phi_{0}=\frac{2\left(\frac{2}{\kappa}(Af)(1-f)
\left(\frac{3(Af)^{2}}{\kappa}\right)
^{\alpha+(1+\alpha)\psi}\right)
^{\frac{1}{2}}}{f+2(f-1)\left[\alpha+(1+\alpha)\psi\right]}
t^{\frac{f+2(f-1)\left[\alpha+(1+\alpha) \psi\right]}{2}},
\end{equation}
where $\phi_{0}$ is an integration constant at $t=0$. Without loss
of generality, we can take $\phi_{0}=0$ to express time in terms of
scalar field as
\begin{equation}\label{9}
t=\left[\frac{\phi\left[f+2(f-1)\left[\alpha+(1+\alpha)
\psi\right]\right]}{2\left(\frac{2}{\kappa}(Af)(1-f)
\left(\frac{3(Af)^{2}}{\kappa}\right)
^{\alpha+(1+\alpha)\psi}\right)
^{\frac{1}{2}}}\right]^{\frac{2}{f+2(f-1)
\left[\alpha+(1+\alpha)\psi\right]}}.
\end{equation}
Using Eq.(\ref{a}), standard scalar potential (\ref{7}) is as
follows
\begin{eqnarray}\nonumber
V(\phi)&=&(1+\omega)\left[\left[\frac{\phi\left[f+2(f-1)
\left[\alpha+(1+\alpha)\psi\right]\right]}{2\left(\frac{2}{\kappa}(Af)(1-f)
\left(\frac{3(Af)^{2}}{\kappa}\right)
^{\alpha+(1+\alpha)\psi}\right)
^{\frac{1}{2}}}\right]^{\frac{4(f-1)(1+\alpha)}{f+2(f-1)
\left[\alpha+(1+\alpha)\psi\right]}}\right.\\
\label{11}&\times&\left(\frac{3(Af)^{2}}{\kappa}\right)
^{1+\alpha}-\left.\frac{A}{1+\omega}\right]^{\psi+1}.
\end{eqnarray}

The slow-roll parameters and number of e-folds are found through
Eq.(\ref{13*}) using Eq.(\ref{9}). Another scalar field $\phi_{1}$
is produced at the beginning of inflation epoch, where $\epsilon=1$.
The standard scalar power spectrum during intermediate era can be
calculated by inserting Eq.(\ref{6}) in (\ref{7*}) and then using
Eq.(\ref{9}), it follows that
\begin{eqnarray}\nonumber
\mathcal{P_{R}}&=&\frac{(Af)^3}{(1-f)}\left(\frac{\kappa}{8\pi^2}\right)
\mu^{\frac{3f-2}{f}}\left[ \mu^{\frac{2(f-1)}{f}}
\left(\frac{3}{\kappa}\right)
(Af)^{2}\right]^{-\alpha-(1+\alpha)\psi}\left[1-\frac{A}{1+\omega}
\right.\\\label{15*}&\times&\left.\mu^{\frac{2(1-f)(1+\alpha)}{f}}\left(\frac{\kappa}{3(Af)^{2}}
\right)^{1+\alpha}\right]^{-\psi},
\end{eqnarray}
where $\mu=\frac{1+f(N-1)}{Af}$. Equations (\ref{8*}) provides
$P_{g}$ and $n_{s}$ as a function of $N$, respectively
\begin{eqnarray}\nonumber
n_{s}-1&=&\frac{2-3f}{Af}\mu^{-1}
+2\left[-\alpha-(1+\alpha)\psi\right]\left(\frac{f-1}{f}\right)
\mu^{\frac{(1-f)}{f}}\\\nonumber&-&\psi
\left[1-\frac{A}{1+\omega}\left(\frac{\kappa}{3(Af)^{2}}
\right)^{1+\alpha}\mu^{\frac{2(1-f)(1+\alpha)}{f}-1}\right]
\left[\frac{2(f-1)(1+\alpha)}{f(1+\omega)}\right.\\\label{17*}&\times&\left.\left(\frac{\kappa}{3(Af)^{2}}
\right)^{1+\alpha}\mu^{\frac{2(1-f)(1+\alpha)}{f}-1}\right], \quad
P_{g}=\left(\frac{2\kappa}{\pi^2}\right)(Af)^2\mu^{\frac{2(f-1)}{f}}.
\end{eqnarray}
\begin{figure}
\centering\epsfig{file=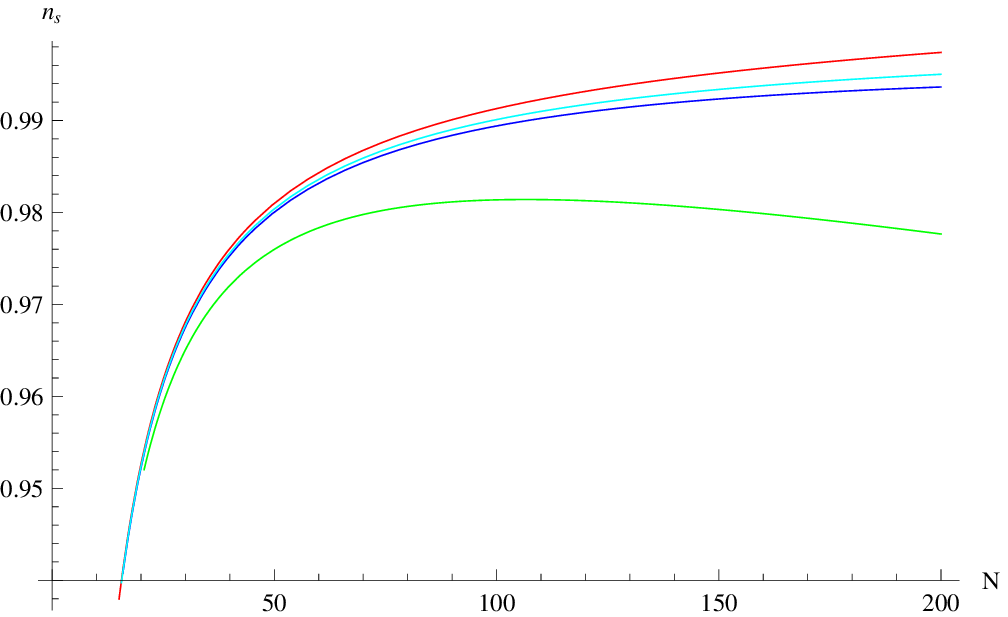,
width=0.45\linewidth}\epsfig{file=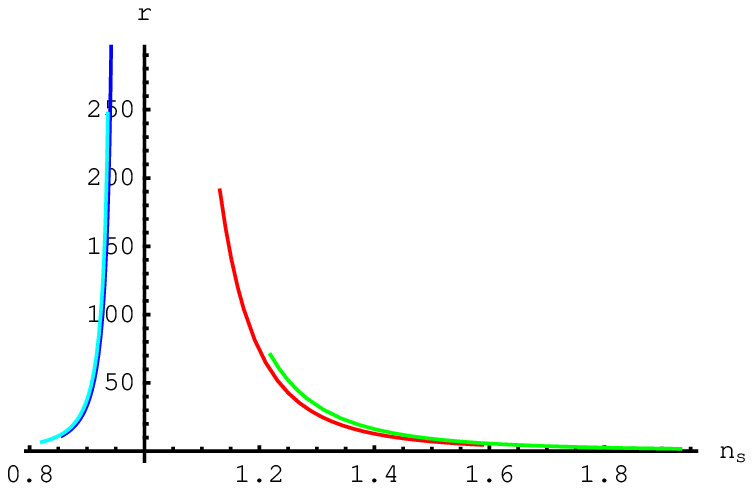,
width=0.45\linewidth}\caption{(left) $n_{s}$ versus $N$, (right)
tensor-scalar ratio versus $n_{s}$ for $A=8.225\times10^2,~\alpha=
0.775,~\omega=-0.8,~g=\frac{1}{2},~k=1$ (red);
$A=2.635\times10^2,~\alpha=0.81,~\omega=-1.5$ (green);
$A=8.407\times10^2,~\alpha=0.85,~\omega=-1.7$ (blue) and
$A=8.407\times10^2,~\alpha=0,~\omega=0$ (zinc) in intermediate
scenario.}
\end{figure}
Using Eqs.(\ref{15*}) and (\ref{17*}), the tensor-scalar ratio has
the form
\begin{eqnarray}\nonumber
r(N)&=&16\left(\frac{1-f}{Af}\right)\mu^{-1}
\left[\left(\frac{3}{\kappa}\right)(Af)^{2}\mu
^{\frac{2(f-1)}{f}}\right]^{\alpha+(1+\alpha)\psi}
\left[1-\frac{A}{1+\omega}\right.
\end{eqnarray}
\begin{eqnarray}
\nonumber&\times&\left.\left(\frac{\kappa}{3(Af)^{2}}
\right)^{1+\alpha}\mu^{\frac{2(1-f)(1+\alpha)}{f}}\right]^{\psi}.
\end{eqnarray}
The left panel of Figure \textbf{1} shows an increasing behavior of
$n_{s}$ with respect to $N$. The observed value of $n_{s}=0.96$
corresponds to $N\approx20$ for all values of parameters which
indicates physical compatibility of these model parameters with
WMAP7 data. The right graph of Figure \textbf{1} shows that red and
green $r-n_{s}$ trajectories are decreasing while other two are
increasing. We see that none of the case is compatible with WMAP7
data as the observed value $0.96\leq n_{s}\leq1$ does not lie in the
region $r\leq0.36$ during intermediate scenario.

\subsubsection{Logamediate Inflation}

Logamediate inflationary era is motivated by imposing weak general
conditions on the indefinite expanding cosmological models. It has
been proved that the power spectrum is either red or blue tilted for
this type of inflation. The scale factor satisfies \cite{26}
\begin{equation}\label{b}
a(t)=a_{0}\exp[A(\ln t)^{\lambda}],\quad\lambda>1.
\end{equation}
For $\lambda=1$, it is converted to power-law inflation. During
logamediate inflation, Eq.(\ref{6}) has the following solution
\begin{eqnarray}\nonumber
\phi(t)-\phi(t_{0})&=&-\Xi(t)\left[\left(\frac{2}{\kappa}\right)
\left(\frac{3}{\kappa}\right)^{\alpha+(1+\alpha)\psi}
(A\lambda)^{1+2(\alpha+(1+\alpha)\psi)}\right]^{\frac{1}{2}}
\\\label{15}&\times&\left[\frac{A}{1+\omega}\left(\frac{\kappa}{3}
\right)^{1+\alpha}(A\lambda)^{-2(1+\alpha)}\right]^{\frac{\psi}{2}},
\end{eqnarray}
where $\Xi(t)=\gamma[\frac{\lambda+2\alpha(\lambda-1)}{2},\alpha\ln
t]$ ($\gamma$ is incomplete gamma function). From the above
equation, $t$ is calculated in terms of $\phi$ as
\begin{eqnarray}\nonumber
t&=&\Xi^{-1}\left[-\phi\left[\left(\frac{2}{\kappa}\right)
\left(\frac{3}{\kappa}\right)^{\alpha+(1+\alpha)\psi}
(A\lambda)^{1+2(\alpha+(1+\alpha)\psi)}\right]^{-\frac{1}{2}}
\right.\\\label{16}&\times&\left.\left[\frac{A}{1+\omega}
\left(\frac{\kappa}{3}\right)^{1+\alpha}
(A\lambda)^{-2(1+\alpha)}\right]^{-\frac{\psi}{2}}\right].
\end{eqnarray}
The corresponding Hubble parameter, standard scalar potential,
slow-roll as well as number of e-folds can be calculated as in the
intermediate case.

The scalar and tensor perturbed parameters in terms of $N$ can be
written as
\begin{eqnarray}\nonumber
\mathcal{P_{R}}&=&\left(\frac{\kappa}{8\pi^2}\right)\left(\frac{\kappa}{3}\right)
^{\alpha+(1+\alpha)\psi}\frac{(A\lambda)
^{3-2(\alpha+(1+\alpha)\psi)}}{(1-\lambda)}
\varepsilon^{\frac{3\lambda-2(\lambda-1)
(\alpha+(1+\alpha)\psi)-2}{\lambda}}\\\nonumber&\times&\exp
\left[\left(\frac{2(\alpha+(1+\alpha)\psi)-1}{\lambda}\right)\varepsilon\right]
\left[1-\frac{A}{1+\omega}\left(\frac{\kappa}{3(A\lambda)^2}\right)^{1+\alpha}
\right.\\\nonumber&\times&\left.\varepsilon^{\frac{-2(1+\alpha)}{\lambda}}\exp
\left[\frac{2\varepsilon(1+\alpha)}{\lambda}\right]\right]^{-\psi},\quad
\mathcal{P}_{g}=\left(\frac{2\kappa}{\pi^2}\right)(A\lambda)^{2}\varepsilon^{\frac{\lambda-1}{\lambda}}\exp
\left[-\frac{\varepsilon}{\lambda}\right],
\end{eqnarray}
where $\varepsilon=\left[\frac{N}{A}+(A\lambda)
^{\frac{\lambda}{1-\lambda}}\right]$. Using $\mathcal{P_{R}}$, we
obtain scalar spectral index
\begin{figure}
\centering\epsfig{file=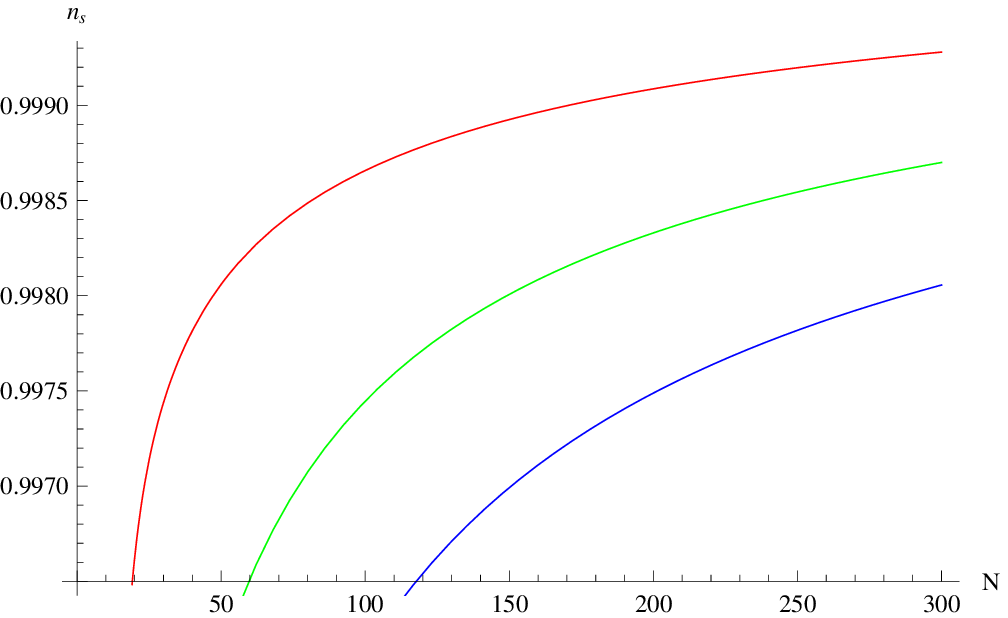,
width=0.45\linewidth}\epsfig{file=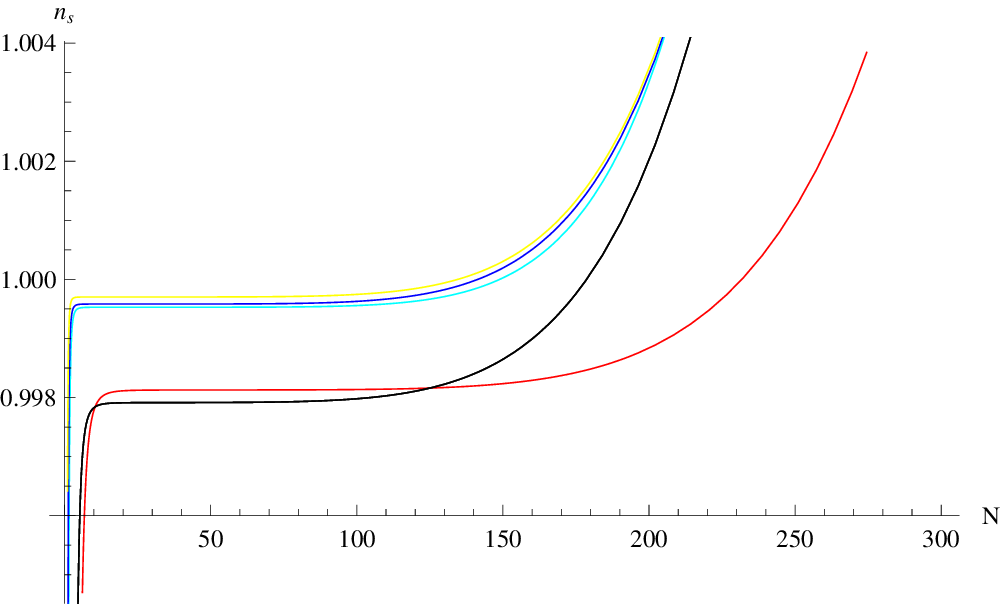,
width=0.45\linewidth}\caption{(left) $n_{s}$ versus $N$ for
$A=8.225\times10^2,~\alpha= 0.775,~\omega=-0.8,~\lambda=10$ (red),
$50$ (green), $70$ (blue), $k=1$ (right) graph for
$A=2.635\times10^2,~\alpha= 0.81,~\omega=-1.5,~\lambda=10$ (red),
$50$ (zinc), $70$ (purple); $A=8.407\times10^2,~\alpha=
0.85,~\omega=-1.7,~\lambda=10$ (black), $50$ (blue), $70$ (yellow)
in logamediate scenario.}
\end{figure}
\begin{eqnarray}\nonumber
n_{s}-1&=&\left(\frac{3\lambda-2(\lambda-1)
(\alpha+(1+\alpha)\psi)-2}
{A\lambda}\right)\varepsilon^{\frac{3\lambda-2(\lambda-1)
(\alpha+(1+\alpha)\psi)-2}{\lambda}}\\
\nonumber&+&\frac{2(\alpha+(1+\alpha)\psi-1)}
{A\lambda}-\psi\left[\frac{A}{1+\omega}
\left(\frac{\kappa}{3}\right)^{1+\alpha}(A\lambda)
^{-2(1+\alpha)}\varepsilon^{\frac{-2(1+\alpha)}{\lambda}}\right.\\\nonumber&\times&\left.\exp
\left[\frac{2\varepsilon(1+\alpha)}{\lambda}\right]\right]
\left[\varepsilon^{-1}-1\right]\left[1-\frac{A}{1+\omega}
\left(\frac{\kappa}{3}\right)^{1+\alpha}(A\lambda)
^{-2(1+\alpha)}\varepsilon^{\frac{-2(1+\alpha)}{\lambda}}\right.\\\nonumber&\times&\left.\exp
\left[\frac{2\varepsilon(1+\alpha)}{\lambda}\right]\right]^{-1}.
\end{eqnarray}
The graphical behavior of $n_{s}$ versus $N$ for different values of
the model parameters is shown in Figure \textbf{2}. The left graph
shows that spectral index is an increasing function of $N$ which
confirms the compatibility of the model with recent observations. In
the right graph, zinc, yellow and blue curves correspond to $N=0$
for $n_{s}\leq1$. Consequently, for all choices of free parameters,
the model remains consistent with WMAP7 data. The tensor-scalar
ratio becomes
\begin{eqnarray}\nonumber
r(N)&=&\frac{16(1-\lambda)}{A\lambda}\left(\frac{\kappa(A\lambda)^2}{3}\right)
^{(\alpha+(1+\alpha)\psi)}\varepsilon^{\frac{1-2\lambda+2(\alpha
+(1+\alpha)\psi)(\lambda-1)}{\lambda}}\exp
\left[\varepsilon\left(\frac{1}{\lambda}-\frac{2}{\lambda}\alpha
\right.\right.\\\nonumber&+&\left.\left.(1+\alpha)\psi\right)\right]\left[1-\frac{A}{1+\omega}
\left(\frac{\kappa}{3}\right)^{1+\alpha}(A\lambda)
^{-2(1+\alpha)}\varepsilon^{\frac{-2(1+\alpha)}{\lambda}}\exp
\left[\frac{2(1+\alpha)}{\lambda}\varepsilon\right]\right]^{\psi}.
\end{eqnarray}
\begin{figure}
\centering\epsfig{file=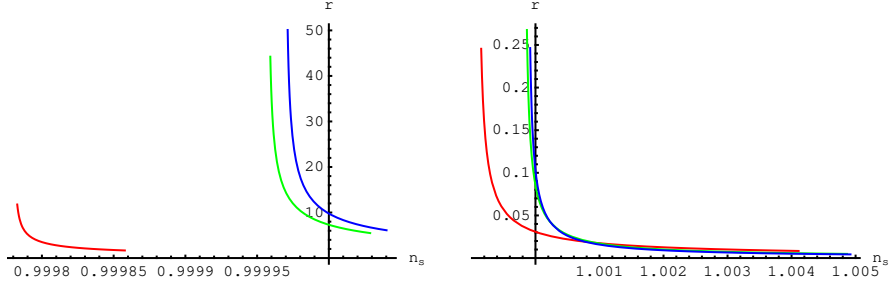,
width=0.45\linewidth}\epsfig{file=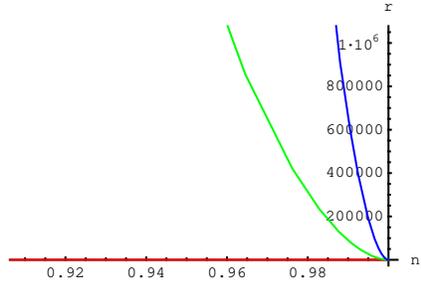,
width=0.45\linewidth}\caption{Tensor-scalar ratio versus $n_{s}$ in
logamediate scenario.}
\end{figure}
During logamediate scenario, Figures \textbf{3} and \textbf{4} show
similar decreasing behavior for all possible choices of the model
parameters. In all cases, we cannot have $n_{s}=0.96$ in the allowed
range of $r\leq0.36$ which is compatible with WMAP7 data.
\begin{figure}
\center\epsfig{file=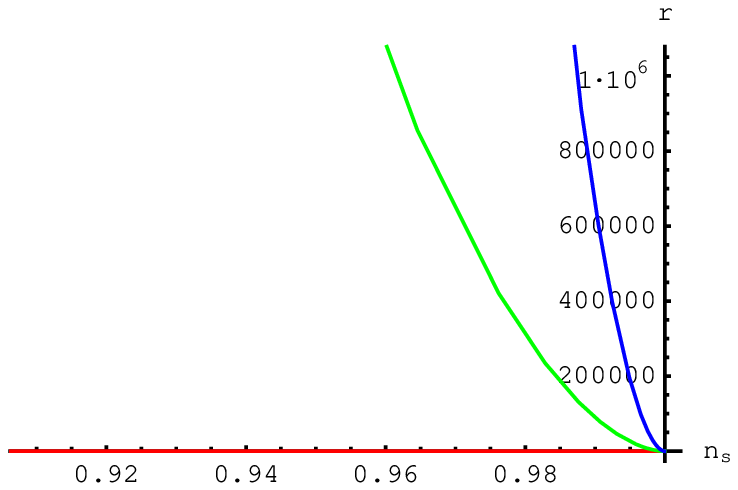,
width=0.45\linewidth}\caption{Tensor-scalar ratio versus $n_{s}$ in
logamediate scenario.}
\end{figure}

\subsection{Tachyon Scalar Field}

In this section, we discuss the intermediate and logamediate
inflationary scenarios in the presence of tachyon scalar field.

\subsubsection{Intermediate Inflation}

The solution of tachyon field during intermediate scenario is given
by Eq.(\ref{3*})
\begin{equation}\nonumber
\phi(t)=\left[\frac{2(1-f)}
{3(Af)(1+\omega)}\right]^{\frac{1}{2}}\left[\frac{A}{(1+\omega)
\left(\frac{3(Af)^2}{\kappa}\right)^{1+\alpha}\chi}\right]t^{\frac{\chi}{2}},
\end{equation}
which gives
\begin{equation}\label{19}
t=\left[\phi\left[\frac{3(Af)(1+\omega)}{2(1-f)}\right]
^{\frac{1}{2}}\left[\frac{(1+\omega)
\left(\frac{3(Af)^2}{\kappa}\right)^{1+\alpha}\chi}{A}\right]
\right]^{\frac{2}{\chi}},
\end{equation}
where $\chi=f+2\alpha(f-1)-4(1+\alpha)(f-1)$. The scalar and tensor
perturbed parameters in terms of $N$ are
\begin{eqnarray}\nonumber
\mathcal{P_{R}}&=&\left(\frac{\kappa}{3}\right)^{\alpha}\left(\frac{\kappa}{8\pi^2}\right)
\frac{(Af)^{3-2\alpha}}{(1-f)}(1+\omega)^{\frac{1}{2}}
\mu^{\frac{(f-1)(3-2\alpha)}{f}}
\left[\left(\frac{3(Af)^{2}}{\kappa}\right)^{(1+\alpha)}
\mu^{\frac{(f-1)(1+\alpha)}{f}} \right.\\
\nonumber&-&\left.\frac{A}{1+\omega}\right]^{-\psi}\left[1+\left(\frac{2}{\kappa}\right)
\left(\frac{3}{\kappa}\right)^{\alpha}(Af)^{2\alpha+1}(f-1)
\mu^{\frac{(f-1)(1+2\alpha)}{f}}
\left[\left(\frac{3(Af)^{2}}{\kappa}
\right)^{1+\alpha}\right.\right.\\\nonumber&\times&\left.\left.\mu^{\frac{(f-1)(1+\alpha)}{f}}
-\frac{A}{1+\omega}\right]^{-1}\right]^{-\frac{1}{2}},\quad
\mathcal{P}_{g}=\left(\frac{2\kappa}{\pi^2}\right)(Af)^2\mu^{\frac{2(f-1)}{f}}.
\end{eqnarray}

The corresponding scalar spectral index is
\begin{eqnarray}\nonumber
n_{s}-1&=&\frac{(1-f)(3-2\alpha)}{Af}\mu^{-1}
-\left(\frac{3}{\kappa}\right)^{1+\alpha}(1+\alpha)(f-1)(Af)^{1+2\alpha}
\mu^{\frac{(f-1)(1+\alpha)}{f}-1}
\\\nonumber&\times&\left[\mu^{\frac{(f-1)(1+\alpha)}{f}}
\left(\frac{3(Af)^2}{\kappa}\right)^{1+\alpha}
-\frac{A}{1+\omega}\right]^{-1}
-\frac{1}{2}\left[1+\left(\frac{2}{\kappa}\right)
\left(\frac{3}{\kappa}\right)^{\alpha}(f-1)\right.\\\nonumber&\times&\left.(Af)^{1+2\alpha}
\mu^{\frac{(f-1)(1+2\alpha)}{f}}
\left[\left(\frac{3(Af)^2}{\kappa}\right)^{1+\alpha}
\mu^{\frac{(f-1)(1+\alpha)}{f}}\frac{A}{1+\omega}\right]^{-1}\right]^{-1}
\end{eqnarray}
\begin{eqnarray}
\nonumber&-&\left[\left(\frac{2}{\kappa}\right)
\left(\frac{3(Af)^2}{\kappa}\right)^{\alpha}(f-1)^2(1+2\alpha)
\mu^{\frac{2\alpha(f-1)-1}{f}}
\left[\left(\frac{3}{\kappa}\right)^{1+\alpha}(Af)^{2(1+\alpha)}
\mu^{\frac{(f-1)(1+\alpha)}{f}}
\right.\right.\\\nonumber&-&\left.\left.\frac{A}{1+\omega}\right]^{-1}\left(\frac{2}{\kappa}\right)
\left(\frac{3}{\kappa}\right)^{1+2\alpha}(Af)^{2(1+2\alpha)}(1-f)^2(1+\alpha)
\mu^{\frac{(f-1)(2+3\alpha)-f}{f}}\right.
\\&\times&\left.\left[\left(\frac{3}{\kappa}\right)^{1+\alpha}(Af)^{2(1+\alpha)}
\mu^{\frac{(f-1)(1+\alpha)}{f}}
-\frac{A}{1+\omega}\right]^{-2}\right].
\end{eqnarray}
\begin{figure}
\centering\epsfig{file=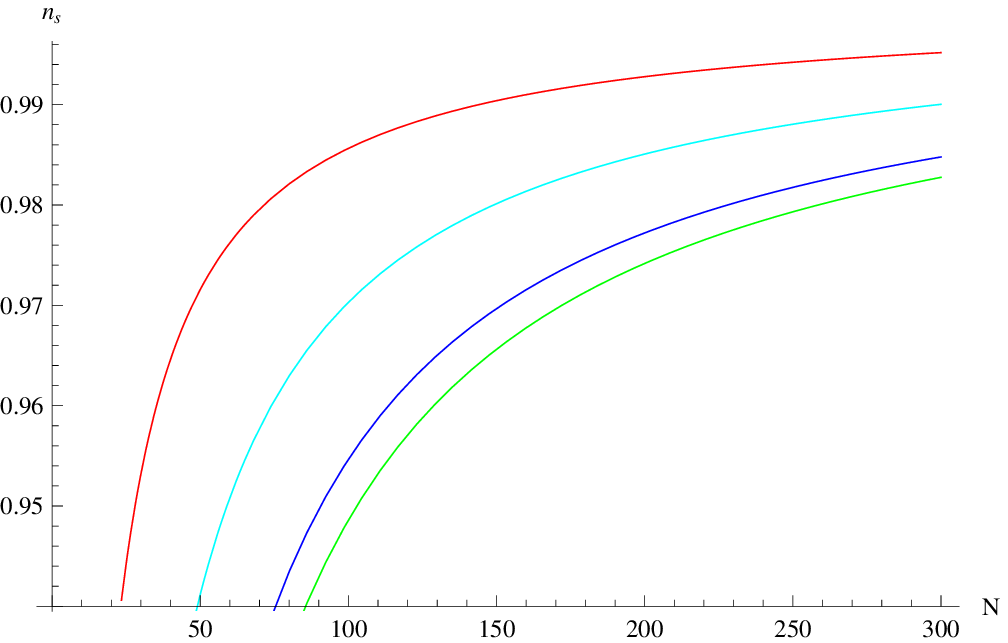,
width=0.45\linewidth}\epsfig{file=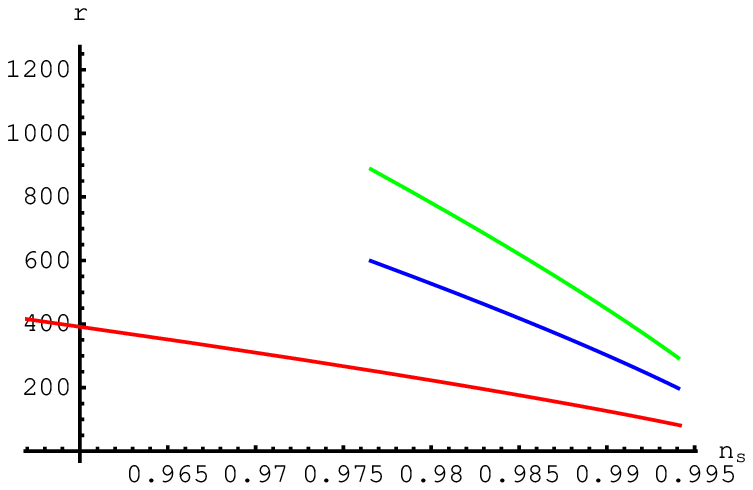,
width=0.45\linewidth}\caption{(left) $n_{s}$ versus $N$ (right)
tensor-scalar ratio versus $n_{s}$ in intermediate scenario.}
\end{figure}
From $\mathcal{P_{R}}$ and $\mathcal{P}_g$, we find tensor-scalar
ratio as
\begin{eqnarray}\nonumber
r(N)&=&16\left(\frac{3}{\kappa}\right)^{\alpha}(1+\omega)
^{-\frac{1}{2}}(1-f)(Af)^{2\alpha-1}
\mu^{\frac{(f-1)(2\alpha-1)}{f}}
\left[\left(\frac{3}{\kappa}\right)^{1+\alpha}(Af)^{2(1+\alpha)}
\right.\\\nonumber&\times&\left.\mu^{\frac{(f-1)(1+\alpha)}{f}}
-\frac{A}{1+\omega}\right]^{\psi}
\left[1+\left(\frac{2}{\kappa}\right)
\left(\frac{3}{\kappa}\right)^{\alpha}(f-1)(Af)^{1+2\alpha}
\mu^{\frac{(f-1)(1+2\alpha)}{f}}\right.
\\\nonumber&\times&\left.\left[\left(\frac{3}{\kappa}\right)^{1+\alpha}(Af)^{2(1+\alpha)}
\mu^{\frac{(f-1)(1+\alpha)}{f}}
-\frac{A}{1+\omega}\right]^{-1}\right]^{\frac{1}{2}}.
\end{eqnarray}
The left graph of Figure \textbf{5} represents increasing behavior
for all four choices of the parameters. In this case, the value of
$n_{s}=0.96$ corresponds to $N\approx20$ (red), $50$ (zinc), $70$
(blue), $90$ (green). Thus the GCCG inflationary intermediate model
with tachyon field is compatible with WMAP7 data. While the right
graph of Figure \textbf{5} shows that the curves in $r-n_{s}$ plane
are decreasing which indicate incompatibility of this model with
recent observations. The physically acceptable range of the
tensor-scalar ratio is not attained at $n_{s}=0.96$ during
intermediate scenario using tachyon field.

\subsubsection{Logamediate Inflation}

Using logamediate scale factor in Eq.(\ref{3*}), we have
\begin{equation}\label{20}
\phi(t)-\phi(t_{0})=\left[\left(\frac{2}{\kappa}\right)
\left(\frac{3}{\kappa}\right)^{\alpha}\frac{\lambda-1}
{1+\omega}(A\lambda)^{1+2\alpha}\right]^{\frac{1}{2}}\Xi(t),
\end{equation}
which provides $t$ in terms of $\phi$ (by assuming $\phi(t_{0})=0$)
as
\begin{equation}\nonumber
t=\Xi^{-1}\left[\left[\left(\frac{2}{\kappa}\right)
\left(\frac{3}{\kappa}\right)^{\alpha}\frac{\lambda-1}
{1+\omega}(A\lambda)^{1+2\alpha}\right]^{-\frac{1}{2}}\phi\right].
\end{equation}
\begin{figure}
\centering\epsfig{file=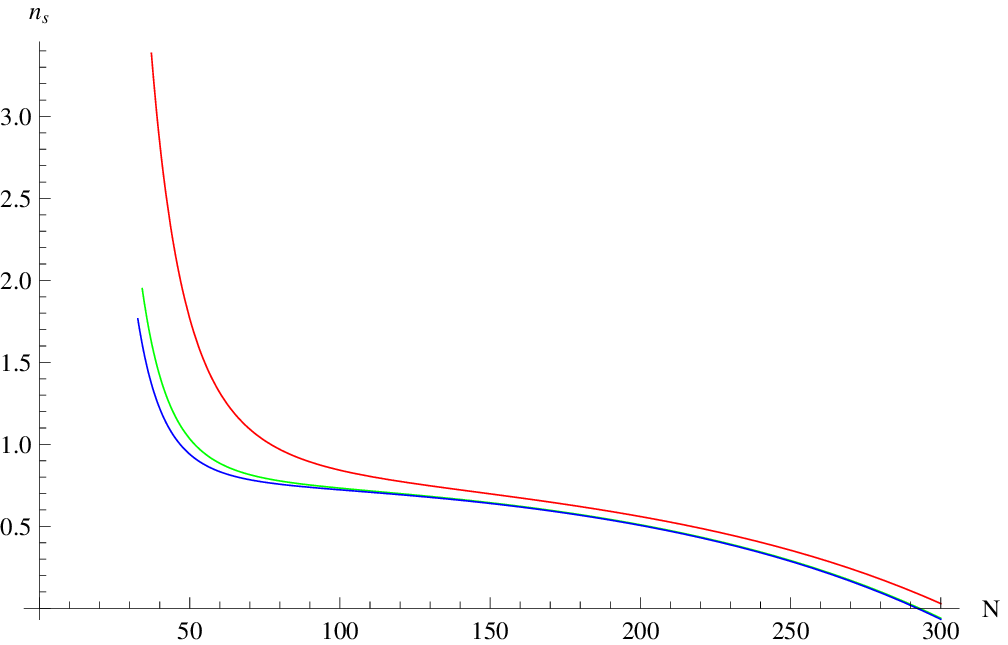,
width=0.45\linewidth}\epsfig{file=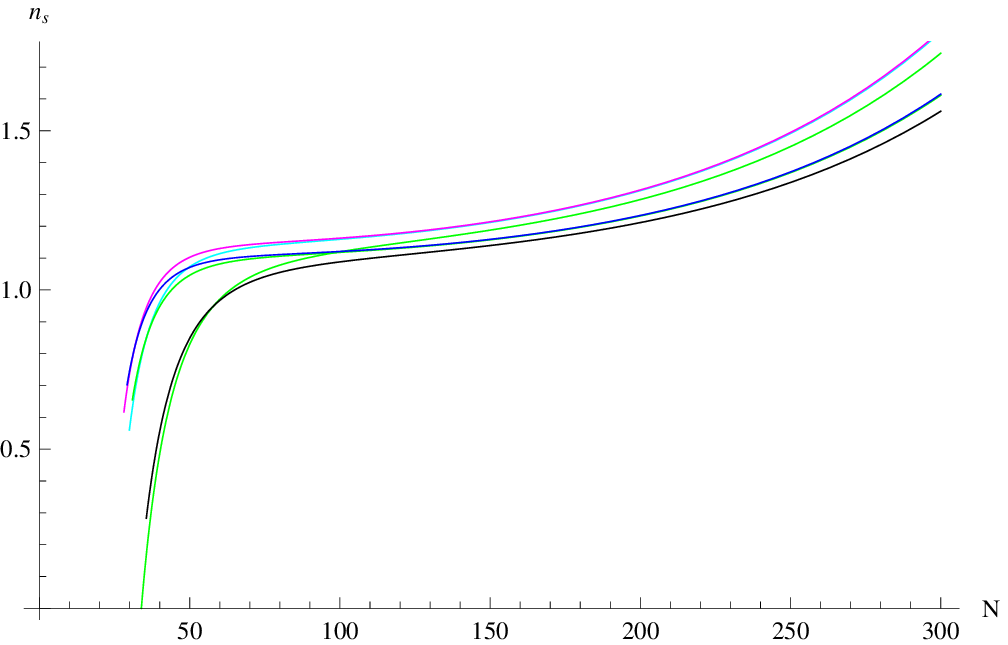,
width=0.45\linewidth}\caption{$n_{s}$ versus $N$ in logamediate
scenario.}
\end{figure}
The scalar as well as tensor power spectra can be expressed as
\begin{eqnarray}\nonumber
\mathcal{P_{R}}&=&\left(\frac{\kappa}{3}\right)^{\alpha}\left(\frac{\kappa}{8\pi^2}\right)
\frac{(1+\omega)^{\frac{1}{2}}}{1-\lambda}(A\lambda)^{3-2\alpha}
\varepsilon^{\frac{3\lambda-2\alpha(\lambda-1)-2}{\lambda}}\exp
\left[\frac{2\varepsilon(\alpha-1)}{\lambda}\right]
\left[\left(\frac{3}{\kappa}\right)^{1+\alpha}\right.\\\nonumber&\times&\left.(A\lambda)
^{2(1+\alpha)}\varepsilon^{\frac{2(1+\alpha)(\lambda-1)}{\lambda}}\exp
\left[\frac{-2\varepsilon(1+\alpha)}{\lambda}\right]-\frac{A}{1+\omega}\right]^{1-\psi},\\\nonumber
\mathcal{P}_{g}&=&\left(\frac{2\kappa}{\pi^2}\right)(A\lambda)^2\varepsilon^{\frac{2(\lambda-1)}{\lambda}}\exp
\left[-\frac{2\varepsilon}{\lambda}\right].
\end{eqnarray}
The scalar spectral index has the form
\begin{eqnarray}\nonumber
n_{s}-1&=&\left(\frac{3\lambda-2\alpha(\lambda-1)-2}{A\lambda}\right)
\varepsilon^{-1}
+\frac{2(\alpha-1)}{A\lambda}+2(1-\psi)(1+\alpha)(\lambda-1)
\left(\frac{3}{\kappa}\right)^{1+\alpha}
\end{eqnarray}
\begin{eqnarray}
\nonumber&\times&(A\lambda)^{2\alpha-1}
\left[\left(\frac{3}{\kappa}\right)^{1+\alpha}(A\lambda)
^{2(1+\alpha)}\varepsilon^{\frac{2(1+\alpha)(\lambda-1)}{\lambda}}\exp
\left[\frac{-2(1+\alpha)}{\lambda}\varepsilon\right]
-\frac{A}{1+\omega}\right]^{-1}
\\\nonumber&\times&\varepsilon^{\frac{2(1+\alpha)(\lambda-1)}
{\lambda}-1}+\frac{1}{1-\lambda}\exp
\left[\frac{-2(1+\alpha)}{\lambda}\left[\frac{N}{A}+(A\lambda)
^{\frac{\lambda}{1-\lambda}}\right]\right].
\end{eqnarray}
\begin{figure}
\centering\epsfig{file=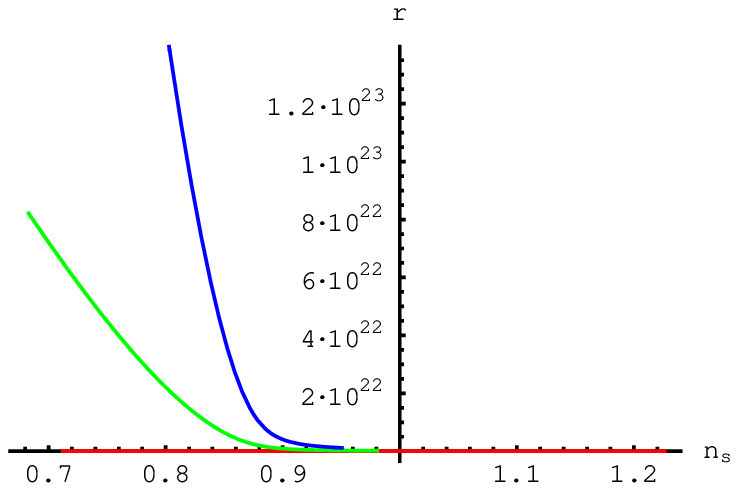,
width=0.45\linewidth}\epsfig{file=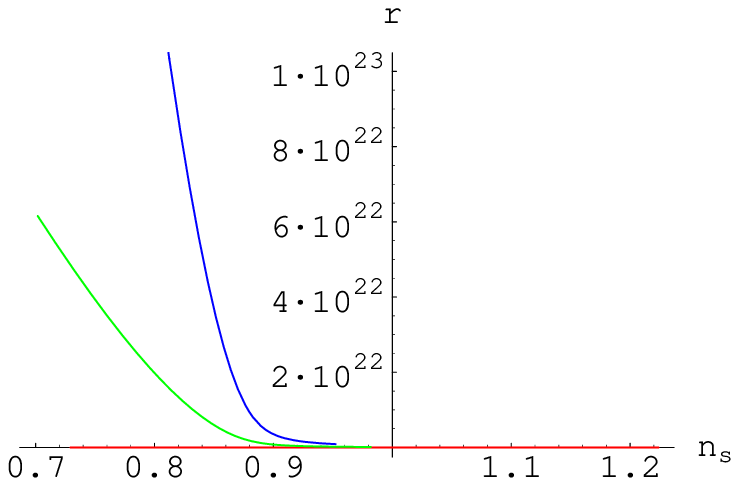,
width=0.45\linewidth}\caption{The left graph of tensor-scalar ratio
versus $n_{s}$ in logamediate scenario.}
\end{figure}
The left and right graphs of Figure \textbf{6} show opposite
behavior to each other for different values of $\lambda$. In the
left graph, when $\lambda$ increases, $n_{s}$ decreases as $N$
increases for all three curves and the constrained $n_{s}=0.96$
corresponds to $N\approx50$ for green and blue curves while
$N\approx100$ for red one. The right graph shows increasing
trajectories and $n_{s}=0.96$ lies in the region $N<50$ for all
choices of the model parameters. The tensor-scalar ratio is
\begin{eqnarray}\nonumber
r(N)&=&-16\left(\frac{3}{\kappa}\right)^{1+\alpha}
\frac{(\lambda-1)}{(\omega+1)^{\frac{1}{2}}}(A\lambda)^{2\alpha-1}
\varepsilon^{\frac{2\alpha(\lambda-1)-\lambda}{\lambda}}\exp
\left[\frac{-2\varepsilon(\alpha+4)}{\lambda}\right]
\\\nonumber&\times&
\left[\left(\frac{3(A\lambda)^2}{\kappa}\right)^{1+\alpha}
\varepsilon^{\frac{2(1+\alpha)(\lambda-1)}{\lambda}}\exp
\left[\frac{-2\varepsilon(1+\alpha)}{\lambda}\right]
-\frac{A}{1+\omega}\right]^{\psi+1}.
\end{eqnarray}
\begin{figure}
\center\epsfig{file=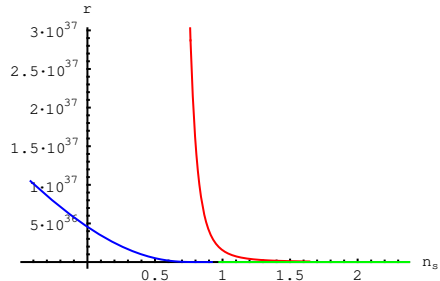,
width=0.45\linewidth}\caption{Tensor-scalar ratio versus $n_{s}$ in
logamediate scenario.}
\end{figure}
Both graphs of Figure \textbf{7} show similar behavior as increasing
$\lambda$ leads to increasing $r-n_{s}$ trajectories. The red curve
in both graphs indicates that $r=0$ for $n_{s}=0.96$ which is
incompatible according to WMAP7 data. Figure \textbf{8} shows
decreasing behavior with increasing $\lambda$. In this case,
$n_{s}=0.96$ corresponds to $r=0$ for $\lambda=50,~70$ while red
curve $(\lambda=10)$ does not lie in the region $r<0.36$.

\section{Concluding Remarks}

In this paper, we have discussed GCCG inflationary universe model
for flat FRW geometry during intermediate as well as logamediate
scenarios. The standard and tachyon scalar fields are considered as
matter content of this universe. In order to study Chaplygin
inspired inflation, we have modified the first Friedmann equation by
applying slow-roll approximation and found solutions of scalar
fields as well as their corresponding potentials. We have also
evaluated slow-roll parameters, number of e-folds, scalar and tensor
power spectra, scalar index and finally the important parameter
tensor-scalar ratio which is constrained by WMAP7 data. We have
checked the physical compatibility of our model with WMAP7 results,
i.e., the standard value $n_{S}=0.96$ must be found in the region
$r<0.36$. The trajectories $N-n_{S}$ and $r-n_{S}$ are plotted to
explore the behavior of these parameters in each case.

By constraining $0.96\leq n_{s}\leq1.002$ and
$\mathcal{P_{R}}(k_{0}=0.002Mpc^{-1})=(2.445\pm0.096)\times10^{-9}$,
according to the observations of WMAP7, we obtain values of the
model parameter as
$A=8.225\times10^2,~2.635\times10^2,~8.407\times10^2$ for $\alpha=
0.775,~0.81,~0.85,~\omega=-0.8,-1.5,-1.7,~g=\frac{1}{2},~\kappa=1$
from Eq.(\ref{15*}). Using these values, we plot the graph of $N$
and $r$ versus $n_{s}$ in intermediate and logamediate scenarios.
The left graph of Figure \textbf{1} shows that $n_{s}=0.96$
corresponds to $N=20$ for all possible choices of the model
parameters during intermediate era. While right panel of Figure
\textbf{1} shows that none of the case is compatible with WMAP7 data
as the observed value $0.96\leq n_{s}\leq1$ does not lie in the
region $r\leq0.36$. The graphical analysis of intermediate era
represents incompatibility of the considered inflationary universe
model for standard scalar field with WMAP7 data. During logamediate
era, the left and right panels of Figure \textbf{2} represent
similar increasing trajectories of $N-n_{s}$ with the increase and
decrease of model parameters $\lambda=10,~50,~70$, respectively.
Thus the number of e-folds remains consistent with observational
value of $n_{s}$ according to WMAP7 data. On the other hand, Figures
\textbf{3} and \textbf{4} show similar decreasing behavior for all
possible choices of the model parameters. The graphical analysis of
this observational parameter of interest $r$ versus $n_{s}$ shows
violation of the observed value of WMAP7 (as $n_{s}=0.96$ does not
correspond to $r\leq0.36$). Thus we conclude that the GCCG
inflationary universe model with a standard scalar field remains
incompatible with observational data of WMAP7.

For tachyon field of the inflationary universe, left plot of Figure
\textbf{5} represents increasing behavior of $N$ with respect to
$n_{s}$ for all four choices of the parameters. In this case, $N$
remains consistent with observational value of $n_{s}$ as for
standard scalar field. While right graph of Figure \textbf{5} shows
incompatibility of this inflationary model with recent observations
of WMAP7 by decreasing $r-n_{s}$ trajectories. In logamediate era,
the left and right graphs of Figure \textbf{6} are opposite in
nature for different values of $\lambda$. In the left panel,
$n_{s}-N$ decreases with the increase of $\lambda$ while right panel
shows increasing trajectories and $n_{s}=0.96$ lies in the region
$N<50$ for all choices of the model parameters. Figures \textbf{7}
and \textbf{8} show similar behavior as obtained for standard scalar
field during logamediate era, i.e., increasing $\lambda$ leads to
decreasing $r-n_{s}$ trajectories. In this case, the red curve in
both graphs matches (i.e., for $n_{s}=0.96,~r=0$) which is not
physical value of $r$ according to WMAP7 data. We conclude that the
inflationary universe model remains incompatible with WMAP7 data for
standard and tachyon scalar fields both in intermediate and
logamediate scenarios. Thus accelerating expansion of the universe
cannot be achieved by using GCCG inflationary universe model in both
intermediate and logamediate regimes.

It is worth mentioning here that all the results for intermediate
regime with standard and tachyon scalar fields reduce to \cite{33}.
Our results for this model support the results of \cite{c} that this
model is less effective as compared to MCG and other DE models.

\vspace{0.25cm}

{\bf Acknowledgment}

\vspace{0.25cm}

We would like to thank the Higher Education Commission, Islamabad,
Pakistan for its financial support through the {\it Indigenous Ph.D.
Fellowship for 5K Scholars, Phase-II, Batch-I}.


\begin{thebibliography}{40}

\bibitem{1} Perlmutter, S. et al.: Nature \textbf{391}(1998)51; Riess, A.G. et
al.: Astron. J. \textbf{116}(1998)1009.

\bibitem{3} Ratra, B. and  Peebles, P.J.E.: Phys. Rev. D
\textbf{37}(1998)3406; Chiba, T., Okabe, T. and Yamaguchi, M.: Phys.
Rev. D \textbf{62}(2000)023511; Padmanabhan, T.: Phys. Rev. D
\textbf{66}(2002)021301; Guo, Z.K. et al.: Phys. Lett. B
\textbf{608}(2005)177.

\bibitem{4} Gorini, V., Kamenshchik, A. and Moschella, U.: Phys. Rev. D
\textbf{67}(2003)063509; Hu, B. and Ling, Y.: Phys. Rev. D
\textbf{73}(2006)123510; Setare, M.R.: J. Cosmol. Astropart. Phys.
\textbf{0701}(2007)023.

\bibitem{5} Zimdahl, W. and Fabris, J.C.: Class. Quantum Grav. \textbf{22}(2005)4311.

\bibitem{6} Bazeia, D.: Phys. Rev. D \textbf{59}(1999)085007.

\bibitem{7} Debnath, U., Banerjee, A. and Chakraborty, S.: Class. Quantum
Grav. \textbf{21}(2004)5609.

\bibitem{34} Gonz\'{a}lez-Diaz, P.F.: Phys. Rev. D \textbf{68}(2003)021303.

\bibitem{a} Kamenshchik, A., Moschella, U. and Pasquier, V.: Phys. Lett. B \textbf{511}(2001)265.

\bibitem{11} Sen, A.: Mod. Phys. Lett. A \textbf{17}(2002)1797.

\bibitem{12} Sami, M., Chingangbam, P. and Qureshi, T.: Phys. Rev. D
\textbf{66}(2002)043530.

\bibitem{13} Gibbons, G.W.: Phys. Lett. B \textbf{537}(2002)1.

\bibitem{14} Bento, M.C., Bertolami, O. and Sen, A.: Phys. Rev. D
\textbf{66}(2002)043507.

\bibitem{15} Dev, A., Alcaniz, J.S. and Jain, D.: Phys. Rev. D
\textbf{67}(2003)023515; Gonz\'{a}lez-Diaz, P.F.: Phys. Lett. B
\textbf{562}(2003)1; Kremer, G.M.: Gen. Relativ. Gravit.
\textbf{35}(2003)1459; Bean, R. and Dore, O.: Phys. Rev. D
\textbf{68}(2003)023515; Chimento, L.P.: Phys. Rev. D \textbf{69}
(2004)123517.

\bibitem{16} Bertolami, O. and Duvvuri, V.: Phys. Lett. B \textbf{640}(2006)121.

\bibitem{17} del Campo, S. and Herrera, R.: Phys. Lett. B \textbf{660}(2008)282;
ibid. \textbf{665}(2008)100.

\bibitem{18} Monerat, G.A. et al.: Phys. Rev. D \textbf{76}(2007)024017.

\bibitem{19} Starobinsky, A.A.: Phys. Lett. B \textbf{91}(1980)99;
Guth, A.: Phys. Rev. D \textbf{23}(1981)347.

\bibitem{20} Gold, B. et al.: Astrophys. J. Suppl.
\textbf{192}(2011)15.

\bibitem{23} Yokoyama, J. and Maeda, K.: Phys. Lett. B \textbf{207}(1988)31.

\bibitem{30} Setare, M.R. and Kamali, V.: Gen. Relativ. Gravit. \textbf{46}(2014)1642.

\bibitem{29} Ferreira, P.G. and Joyce, M.: Phys. Rev. D
\textbf{58}(1998)023503; Barrow, J.D. and Nunes, N.J.: Phys. Rev. D
\textbf{76}(2007)043501.

\bibitem{32} Setare, M.R. and Kamali, V.: arXiv:1309.2452.

\bibitem{32*} Sharif, M. and Saleem, R.: Eur. Phys. J. C \textbf{74}(2014)2738.

\bibitem{33} Herrera, R., Olivares, M. and Videla, N.: Eur. Phys. J. C \textbf{73}(2013)2295.

\bibitem{25} Barrow, J.D. and Liddle, A.R.: Phys. Rev. D \textbf{47}(1993)5219; Starobin-
sky, A.A.: J. Exp. Theor. Phys. Lett. \textbf{82}(2005)169; del
Campo, S., Her- rera, R., Saavedra, J., Campuzano, C. and Rojas, E.:
Phys. Rev. D \textbf{80}(2009)123531; Herrera, R. and Videla, N.:
Eur. Phys. J. C \textbf{67}(2010)499.

\bibitem{36} Martin, J. and Schwarz, D.J.: Phys. Rev. D
\textbf{57}(1998)3302.

\bibitem{37} Zarrouki, R. and Bennai, M.: Phys. Rev. D \textbf{82}(2010)123506.

\bibitem{38} Liddle, A. and Lyth, D.: \textit{Cosmological Inflation and Large-Scale
Structure}, (Cambridge University Press, 2000).

\bibitem{39} Garousi, M.R., Sami, M., Tsujikawa, S.: Phys. Rev. D
\textbf{70}(2004)043536.

\bibitem{24} Sanyal, A.K.: Phys. Lett. B \textbf{645}(2007)1.

\bibitem{26} Barrow, J.D.: Class. Quantum Grav. \textbf{13}(1996)2965.

\bibitem{c} Rudra, P.: Astrophys. Space Sci. \textbf{342}(2012)579.

\end{thebibliography}
\end{document}